\documentclass[aip, jcp, jmp, amsmath,amssymb, preprint]{revtex4-1}
\usepackage{amsmath}
\usepackage{latexsym}
\usepackage{graphicx}
\usepackage{color}
\usepackage{fancyhdr}
\usepackage{hyperref}

\begin{document}
\title{Novel High-Speed Polarization Source For Decoy-State BB84
Quantum Key Distribution Over Free Space and Satellite Links}

\author{Zhizhong Yan}
\email{zyan@iqc.ca} \affiliation{Institute for Quantum Computing,
University of Waterloo, 200 University Avenue W, Waterloo N2L 3G1,
Canada}

\author{Evan~Meyer-Scott}
\affiliation{Institute for Quantum Computing, University of
Waterloo, 200 University Avenue W, Waterloo N2L 3G1, Canada}

\author{Jean-Philippe Bourgoin}
\affiliation{Institute for Quantum Computing, University of
Waterloo, 200 University Avenue W, Waterloo N2L 3G1, Canada}

\author{Brendon L Higgins}
\affiliation{Institute for Quantum Computing, University of
Waterloo, 200 University Avenue W, Waterloo N2L 3G1, Canada}

\author{Nikolay Gigov}
\affiliation{Institute for Quantum Computing, University of
Waterloo, 200 University Avenue W, Waterloo N2L 3G1, Canada}

\author{Allison MacDonald}
\affiliation{Institute for Quantum Computing, University of
Waterloo, 200 University Avenue W, Waterloo N2L 3G1, Canada}
\affiliation{Physics Department, University of Alberta, Edmonton,
Alberta, Canada}

\author{Hannes H\"{u}bel}
\affiliation{Institute for Quantum Computing, University of
Waterloo, 200 University Avenue W, Waterloo N2L 3G1, Canada}
\affiliation{Department of Physics, Stockholm University, Stockholm,
Sweden}

\author{Thomas~Jennewein}\email{thomas.jennewein@uwaterloo.ca}
\affiliation{Institute for Quantum Computing, University of
Waterloo, 200 University Avenue W, Waterloo N2L 3G1, Canada}

\begin{abstract}
To implement the BB84 decoy-state quantum key distribution (QKD) protocol over a lossy ground-satellite
quantum uplink requires a source that has high repetition rate of short laser pulses, long term stability,
and no phase correlations between pulses. We present a new type of telecom optical polarization and
amplitude modulator, based on a balanced Mach-Zehnder interferometer configuration, coupled to a
polarization-preserving sum-frequency generation (SFG) optical setup, generating 532~nm photons with
modulated polarization and amplitude states. The weak coherent pulses produced by SFG meet the challenging
requirements for long range QKD, featuring a high clock rate of 76~MHz, pico-second pulse width, phase
randomization, and 98\% polarization visibility for all states. Successful QKD has been demonstrated using
this apparatus with full system stability up to 160 minutes and channel losses as high
57~dB~\cite{Meyer-Scott_PRA_11}. We present the design and simulation of the hardware through the Mueller
matrix and Stokes vector relations, together with an experimental implementation working in the telecom
wavelength band. We show the utility of the complete system by performing high loss QKD simulations, and
confirm that our modulator fulfills the expected performance.
\end{abstract}

\maketitle

\section{Introduction}
Utilizing ground-to-space quantum communications with satellites to
achieve long-distance quantum key distribution (QKD) has been
theoretically studied and experimentally proven
feasible~\cite{Rarity_NJP_02, Miao_PRA_07_The-feas,
Villoresi_NJP_08, Bonato_NJP_09}. Ultimately, a global quantum
network can be developed by using satellites as trusted nodes. An
satellite uplink scheme is appealing because the energy-demanding
and relatively complex photon source remains at the ground, easing
the requirements on the satellite and granting the ability to use a
variety of different sources. Practical QKD
apparatuses~\cite{Gisin_RevModPhys_03} use either weak coherent
pulse (WCP) sources~\cite{Dixon_OptExp_08, Dixon_10, Jofre_JLT_10,
Jofre_Opt_11, Stucki_NJP250k}, or sources of entangled photon
pairs~\cite{Jennewein_PRL_00, Tanzilli_PPLN_02, meyer-scott:031117,
Ma_PRA_07, Erven_NJP_09, Aspelmeyer_JSTQE_2003_Long-dis}. The major
challenge of an uplink approach is the additional loss stemming from
atmospheric turbulence, leading to low signal compared to the noise
of detector dark counts and stray light.

To perform QKD under such challenging conditions, it is desirable to
have the source emit single photons within the shortest time window
possible. This can be achieved with a mode-locked laser, which
intrinsically provides short pulses and a high repetition rate.
However such a laser has the inevitable problem of possessing phase
correlations between consecutive pulses. This phase correlation
violates an assumption of QKD security
proofs~\cite{Gottesman2004Security,Jofre_JLT_10}, and is unsuitable
as a photon source for QKD without sufficient precaution. To solve
this problem, our source produces green photons at 532~nm wavelength
by the sum-frequency generation (SFG) process, pumped by a pulsed
810~nm laser and a continuous wave (CW) laser at 1550~nm. The hybrid
nature of this design enables us to exploit the repetition rate and
pulse width of the mode-locked Ti:Sapphire laser (810~nm), while
phase randomization is accomplished with the short coherence length
of the telecom laser.  Importantly, the photons at 532~nm allow the
use of detectors with the highest detection figure of
merit~\cite{Hadfield_NatPhoton_09} available. Additionally, we gain
access to fast and stable modulation components designed for the
telecom-band around 1550~nm. In this work, we mainly present the
telecom intensity and polarization modulator (IPM) to encode quantum
keys, as well as the QKD results to verify its performance as one of
the building blocks. The SFG nonlinear optical setup is discussed
elsewhere~\cite{Meyer-Scott_PRA_11,Meyer-Scott_thesis_master}.

The desired performance for satellite-based QKD previously discussed
in, e.g., Refs.~\cite{Jofre_JLT_10} and~\cite{Dios_AppOpt_04},
requires that we consider the following
factors when implementing BB84 QKD protocol \cite{BB84} with the IPM: (1)
Polarization encoding in the four non-orthogonal BB84
states~\cite{Liu_08}. The main advantage of polarization encoding is
that Earth's atmosphere preserves polarization with high
fidelity~\cite{Toyoshima:09}. (2) Compatible with the SFG setup to
produce 532~nm green photons. (3) Compatible with Ti:Sapphire
repetition rate, and readily upgradable to several GHz repetition
rate.

Previous implementations of polarization modulation for QKD include
multiple laser diodes \cite{Liu_OptExpr_10}, a single laser diode
with multiple optical amplifiers \cite{Jofre_Opt_11}, a single phase
modulator with Faraday mirror \cite{Lucio-Martinez_NJP_09}, as well
as a single polarization modulator \cite{Jofre_JLT_10}. The first
two approaches \cite{Liu_OptExpr_10, Jofre_Opt_11}, due to their
using distinct sources of light, face the difficulty of making all
output quantum states identical in frequency, bandwidth, and
intensity; distinguishing information leads to security loopholes.
The latter two methods \cite{Lucio-Martinez_NJP_09, Jofre_JLT_10}
will suffer polarization mode dispersion when two orthogonal modes
have to propagate along unsymmetrical optical axes of the same
modulator crystal or polarization maintaining fiber, leading to poor
polarization states. Long-term thermal stability is also problematic
for these implementations, and all exhibit pulse lengths at least
two orders of magnitude longer than those obtained with a
mode-locked laser.

To avoid the above problems, we present our high speed polarization
modulator in the balanced Mach-Zehnder Interferometer (MZI)
configuration, including detailed descriptions of its design
principles, and analysis of the experimental performance of the
modulator. The paper is
organized as follows: first we present an overview of the polarization and
intensity modulator design, and its role in the entire QKD system;
then we discuss the mathematical modelling of the telecom
polarization and intensity modulator; finally, we show  experimental
results and the expected performance of
decoy-state QKD based on our design.
\section{All-Fiber Telecom Band Intensity and Polarization Modulation System: Design, Modelling and Characterization}
\subsection{Overview of the QKD source design}
Implementing the BB84 protocol with decoy states requires that the
output photons have at least two levels of average photon number. In
the simplest case, this entails a \emph{signal} state with average
photon number $\mu$ and a \emph{decoy} state with average photon
number $\nu$. For both levels, the modulator should output one of
four polarization states (horizontal, vertical, diagonal, or
anti-diagonal), chosen randomly. For all states, the average photon
number per pulse is less than 1 to keep the multi-photon probability
low~\cite{Ma_PRA_05}. However, the decoy-sate QKD protocol allows
for a much larger average photon number (e.g.\ $\mu=0.5$) for signal
pulses, as compared to WCP QKD protocols not employing decoy
states~\cite{Lo_PRL_05}.

\begin{figure}
\centering
\includegraphics[width=0.5\textwidth]{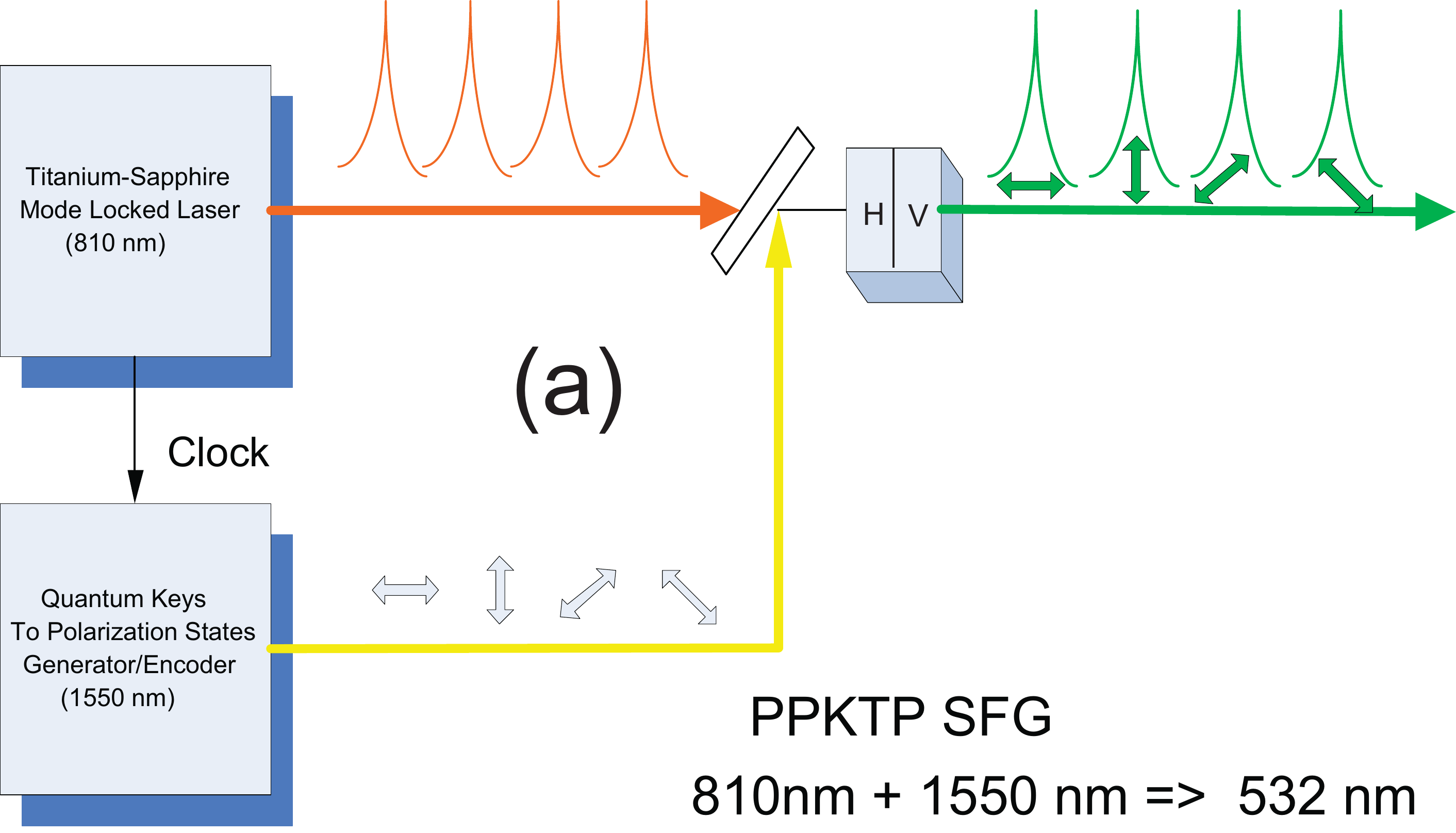}
\includegraphics[width=0.25\textwidth]{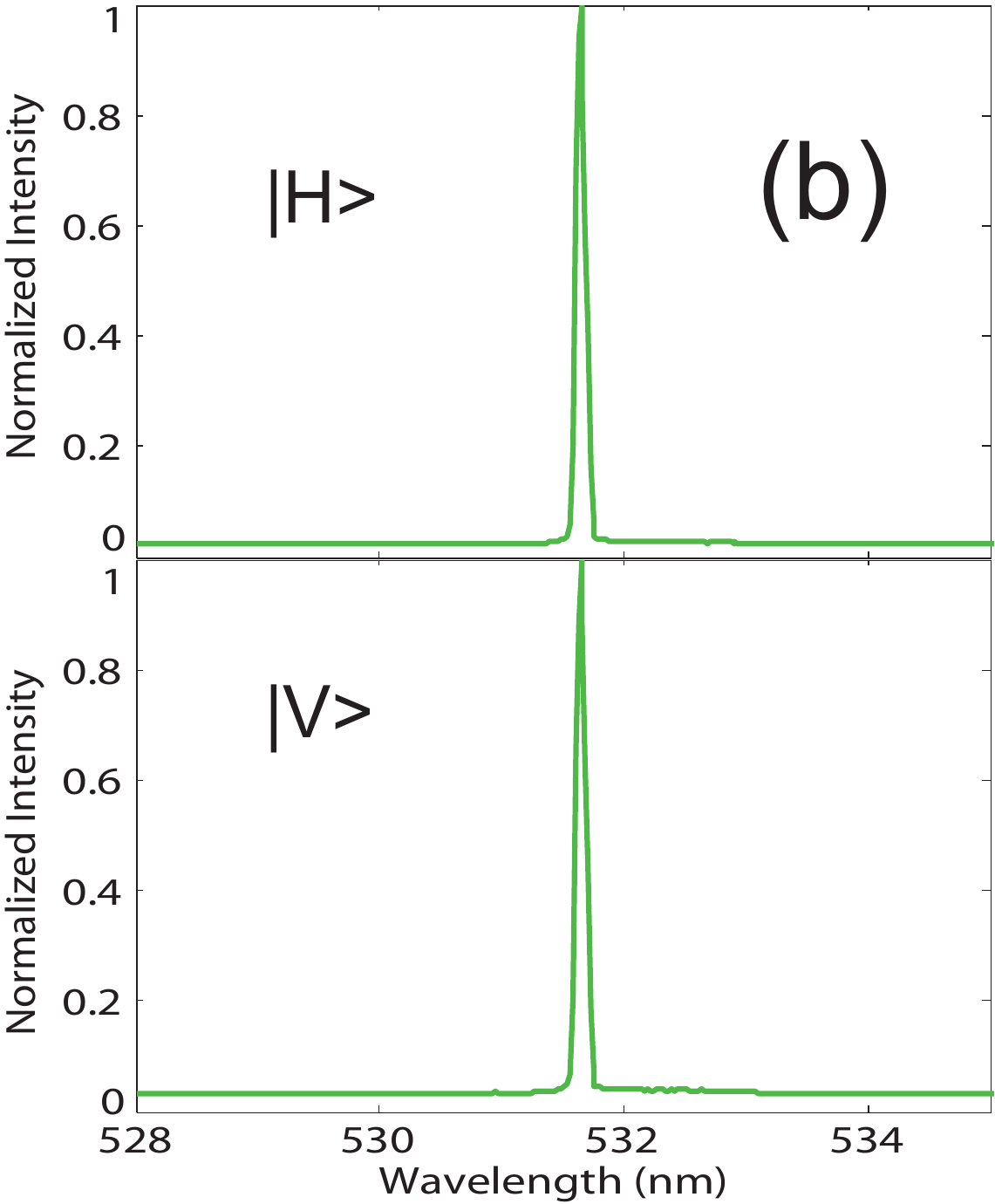}
\caption{(a) Generation of QKD polarization states at 532~nm by polarization-preserving
SFG through a walk-off compensated PPKTP crystal pair (labeled as H
and V, respectively). Light at 810~nm produced by a Ti:Sapphire
laser, mode-locked with 76~MHz repetition rate, is combined with
1550~nm light with polarization state engineered by the IPM synchronized to the repetition rate of the Ti:Sapphire
laser. The SFG green (532~nm) photons combine properties from both
pumping sources, including pulse shape and repetition rate from the
Ti:Sapphire laser, and average photon number per pulse and
polarization states from the modulated telecom laser. (b) The
measured spectrum of green photons produced by our
polarization-preserving SFG optical setup. Horizontally and
vertically polarized photons have identical
spectra.}\label{fig:system_top_architecture}
\end{figure}

Our QKD system is based on the SFG photon scheme as showed in Fig.
\ref{fig:system_top_architecture}, in order to achieve the required
modulation performance, pulse length, and phase randomization at the
desired wavelength. The scheme includes a 1550~nm CW laser modulated
in amplitude and polarization before being converted to 532~nm through SFG with
strong 810~nm pulsed light from the mode-locked Ti:Sapphire laser.
An overview of the SFG solution is illustrated in
Fig.~\ref{fig:system_top_architecture} (a). The SFG green photons'
field intensity $I_3$ for Type-I PPKTP is proportional to the
product $I_1 I_2$  where  $I_1$ and $I_2$ represent the telecom and
810~nm optical power intensities~\cite{boyd}. The linear dependence of the output intensity of green photons (whose spectrum is shown in
Fig.~\ref{fig:system_top_architecture} (b))
on the input intensities of both the telecom and 810~nm pumps
allows direct modulation of the output light through modulation of
only the telecom input. Since we are on at the regime of weak
conversion, the spectrum of the generated photons will not be
perturbed by the intensities of $I_1$ and $I_2$.

The QKD system has two basic parameters that determine the final key
rate and maximum distance: quantum bit error rate (QBER) and sifted
key rate. A nonzero QBER is mainly caused by detector dark counts
and polarization state visibility~\cite{Gisin_RevModPhys_03}. To
minimize QBER, by utilizing temporal filtering of detector and
background noise, we incorporated thin-silicon SPAD detectors in
Bob's receiver. The efficiency of these detectors peaks at a
wavelength around 532~nm.

The sifted key rate is determined by the repetition rate of the
source, the average photon number $\mu$, and channel losses. Pockels
cells are capable of switching at such short wavelengths, but they
require driving electronics operating at a few kilovolts for a $\pi$
phase shift ($V_\pi$), preventing such modulators from being
operated at the repetition rate of a Ti:Sapphire laser
(${\sim}76$~MHz), let alone approaching GHz rates desirable of
high-speed QKD. Instead, our scheme in Fig.
\ref{fig:system_top_architecture} allows for using conventional
electro-optical (EO) waveguide modulators to match the desired
repetition rate.

\subsection{Overview of the telecom all-fiber modulation system}
\begin{figure}
\centering
\includegraphics[width=0.55\textwidth]{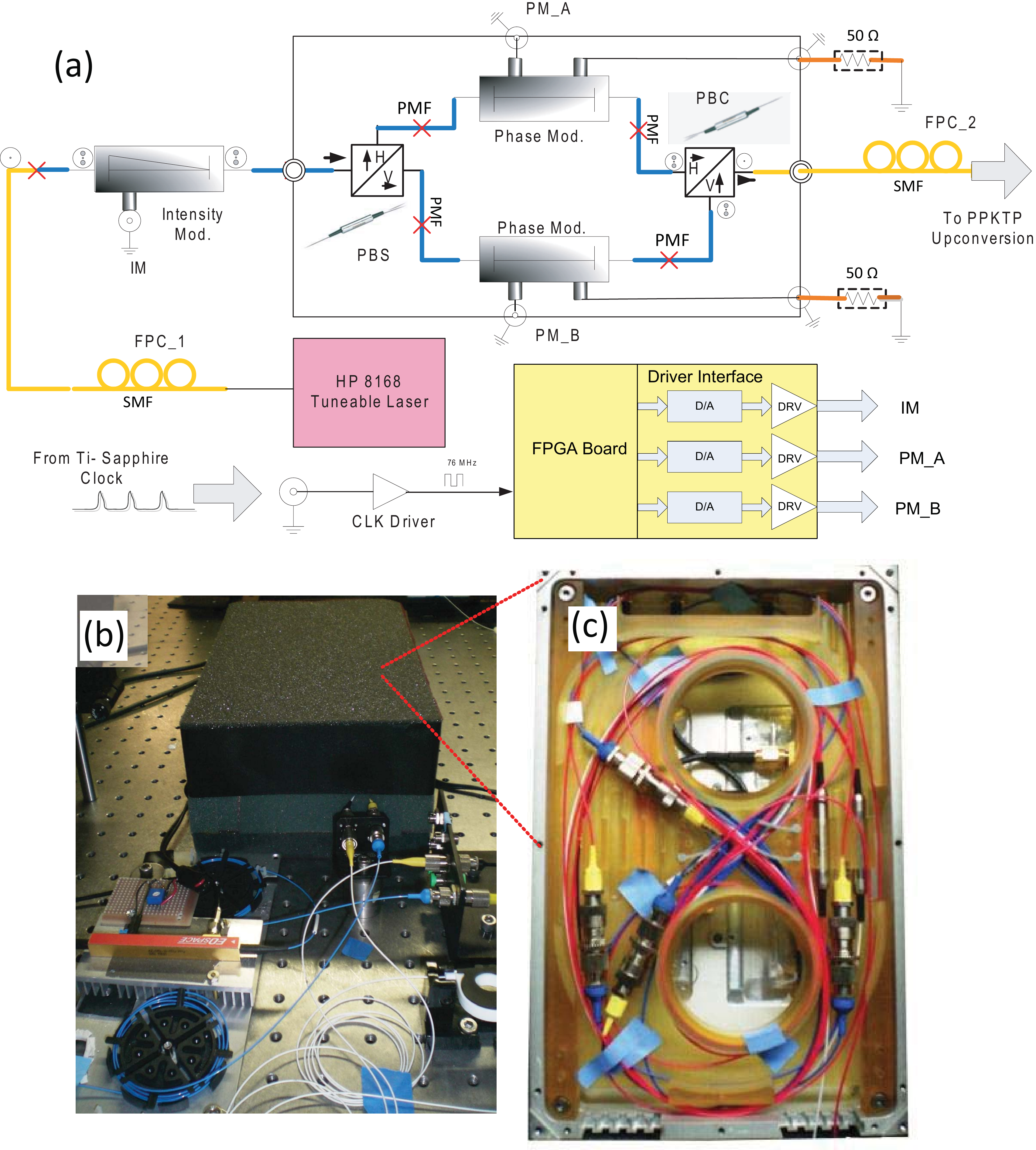}
\caption{(a) Diagram of the intensity and polarization modulator.
Signal and decoy states are realized by the intensity modulator, and
polarization modulation by the balanced MZI, where each arm contains
an EO phase modulator. The input and output to the modulator are a
polarization beam splitter and a polarization beam combiner,
respectively. The FPGA logic and digital \& analog mixed signal
board provide the encoding and driving voltages to the phase
modulators and intensity modulator. The FPGA board is interfaced to
three independent channels of fast digital to analog converter (D/A)
circuits. The RF driving amplifier (DRV) are DC coupled to three
ports of EO modulators. Two paddle-wheel fiber polarization
controllers (FPCs) are used to adjust the input and output
polarization states. Thick yellow lines (color) illustrate standard
single mode fiber (SMF); thick blue lines (color) represent
polarization-maintaining fiber (PMF); each red cross (color)
indicates narrow key FC/PC connectors, enabling high-precision
polarization coupling between two connecting optical fibers. The
FPGA system clock is externally obtained from the PIN diode pulses
of a mode-locked Ti-Sapphire laser. Those pulses are amplified via a
pulse shaping circuit to generate 76 MHz TTL clock signal. (b) Photo
of modulators including intensity (external PM fiber coupled) and
phase modulators, as well as the polarization beam splitters and
combiners (sealed in a thermally insulating box). (c) The core of
this polarization modulator, comprised of a pair of phase modulators
at the lower level, and one each of polarization beam splitter and
combiner. } \label{fig:system_architecture}
\end{figure}

The telecom modulation system for implementing decoy-state BB84
protocol consists of an intensity modulator followed by a
polarization modulator, as realized by two phase modulators in the
balanced MZI configuration that is illustrated in
Fig.~\ref{fig:system_architecture} (a). The physical apparatus of
this scheme is pictured in Fig.~\ref{fig:system_architecture} (b).
The phase and intensity modulators are customized EO LiNbO$_x$
modulators from EOSpace, featuring low insertion loss (${<}2$~dB)
and external 50~$\Omega$ termination of the radio frequency (RF)
driving voltage. Their $V_\pi$ is as low as only a few volts,
compatible with regular analog-digital mixed signal electronics.

To ensure enough key is being received even when channel loss is
high, the system should be capable of modulation at a
few hundred MHz. The factors that limit the system clock rate are
the finite bandwidth of the modulator drivers, the finite speed of
field programmable gate array (FPGA) circuit board, and the
interface circuits. The detailed design is displayed in Fig.
\ref{fig:system_architecture} (a).

The modulator driver circuit is comprised of digital-to-analog
interface to bridge FPGA board and the power amplifier that are both
adapted to be DC coupled to the RF ports of all modulators. To
accomplish wide-band uniformity of the modulation, the driving
signal bandwidth ranges from zero to several hundreds of MHz,
resulting in reliable switching between any two polarization states.
This is particularly important when the RF driving is in the lower
frequency regime (e.g.\ during source testing), because the
resulting heat accumulation is detrimental to the stability of
polarization states, leading to poor QBER performance.
\subsection{Modelling of the telecom all-fiber modulation system}
We model the action of the modulation system using the Jones matrix
approach assuming fully polarized input light and define the
following Jones vector for the horizontally polarized state:
\begin{equation} \label{Jones_in}
|\mathrm{H} \rangle=
\begin{bmatrix}
 1 \\
 0
\end{bmatrix}.
\end{equation}
The system is comprised of an intensity modulator (IM), followed by
a polarization beam splitter (PBS) with its fast axis rotated
$45^\circ$ from the input fast axis (represented by a $\pi/4$
polarization rotation). Both input and output ports of the PBS are
coupled to PM fiber. Following this is the balanced MZI, and after
which is the polarization beam combiner (PBC) with PM input and
single mode (SM) output. The Jones calculus is employed to integrate
all of the optical components into a single matrix $[J]$
\cite{Fujiwara2007Spectros}
\begin{equation}\label{eq_1}
[J] = \left[\mathrm{QWP}\left( - \frac{\pi}{4} \right)\right] \left[
\mathrm{PBC} \right] \left[ \mathrm{MZI} \right] \left[ \mathrm{PBS}
\right] \left[\mathrm{Ro}\left( \frac{\pi}{4} + \delta\right) \right
].
\end{equation}
The last $\left[\mathrm{QWP}\left(- \frac{\pi}{4} \right)\right]$ is
used to rotate from modulation in the circular-diagonal polarization
plane to the linear polarization plane, and is performed by FPC\_2
in the apparatus.

We consider the intensity modulator separately as it acts globally
on the total output intensity independent of polarization. Its
mathematical model, as usually defined, is $ \mathrm{IM}\left( {V_0
} \right)= \frac{1}{2}\left\{ {1 + b\cos \left( \frac{V_0
}{V_\pi}\pi + \Phi _1 \right)} \right\}$ where $0 \leq b \leq 1$ is
modulation depth and $\Phi _1$ is the zero voltage phase, adjustable
by applying an external DC bias. The applied voltage is $V_0$ and
the voltage parameter $V_\pi$ is defined in Ref.~\cite{Li_JLT_03}
as: $ V_\pi   = \frac{\lambda}{n_0^3 r_{ij} }\frac{d}{\gamma L} $ in
which $r_{ij}$ is the linear electro-optical tensor element, $n_0$
is the zero voltage refractive index, $\gamma$ is the optical
confinement coefficient, $L$ is the length of waveguide, and $d$ is
the gap distance of transmission line that is excited by the
modulator electrode input voltage. For the intensity modulator, the
theoretical curve with $b=1$ and $V_\pi=4.0$~V is the best fit for
the experimental verification, and we measured an extinction ratio
of our apparatus reaching 30~dB.

\begin{figure}
\centering
  \includegraphics[width=0.45\textwidth]{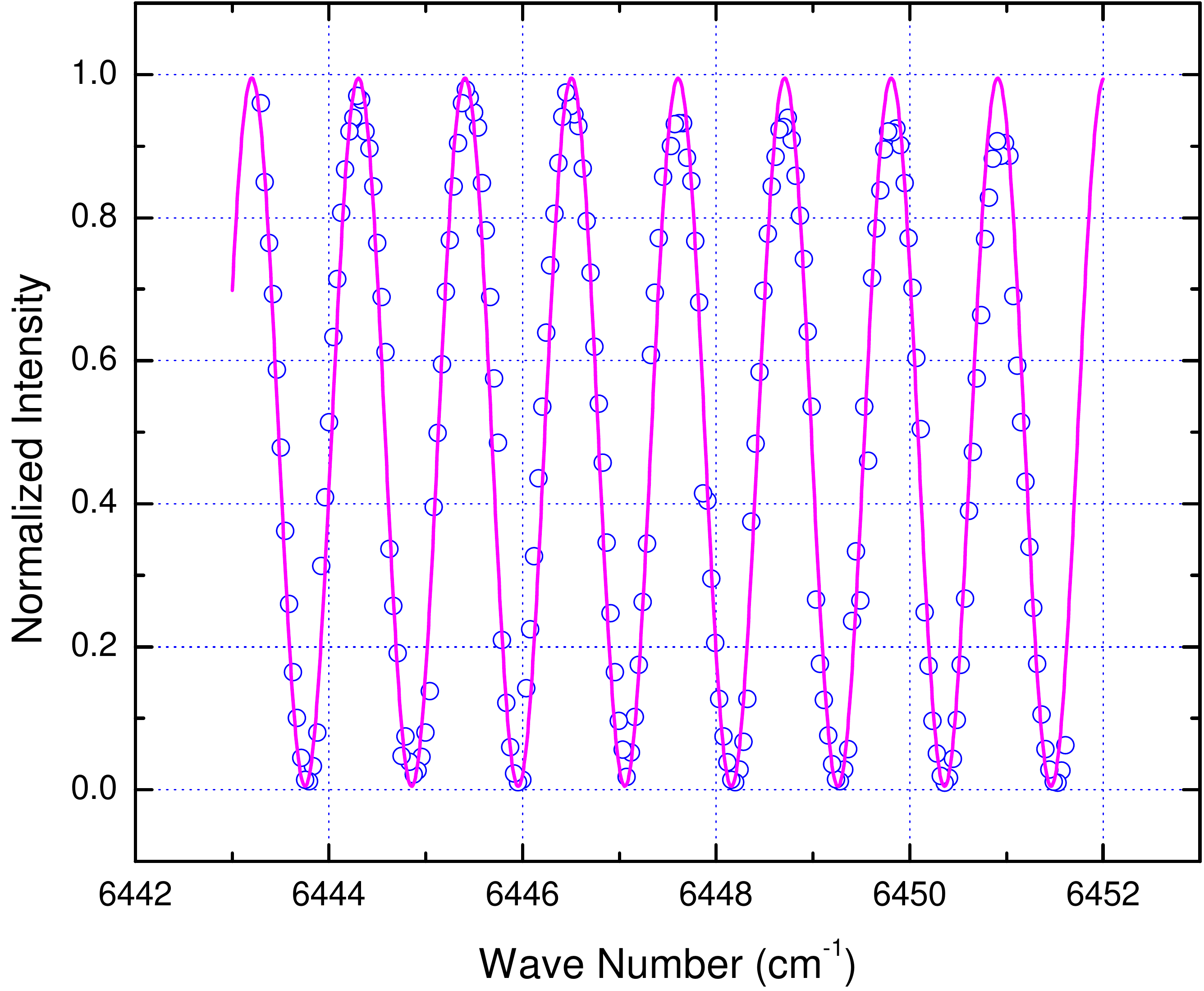}\\
\caption{The measured intensity oscillation after a linear polarizer
and a quarter wave plate when the input wavelength of our MZI
modulator is varied. The solid line represents simulation data;
hollow circles are the measured results.}\label{fig:delta_L}
\end{figure}

In coupling from the IM to the MZI, the actual PBS component has a small rotation offset $\delta$ from the
desired $45^\circ$, resulting in unequal splitting of the input beam. Thus the real device can be modelled
as
\begin{equation}
\mathrm{Ro}\left( {\frac{\pi}{4} + \delta}\right) = \frac{1}{\sqrt
2}
\begin{bmatrix}
   {\cos \delta  - \sin \delta } & {\cos \delta  + \sin \delta }  \\
   { - \cos \delta  - \sin \delta } & {\cos \delta  - \sin \delta }
 \end{bmatrix}.
\end{equation}
In the MZI polarization modulator, the relative phase delay between
the two arms is controlled by high frequency or DC voltages on each
modulator of $V_1$ and $V_2$ respectively, in the push-and-pull
mode. The effect of the PBS and PBC are included implicitly in the
Jones calculus, as the matrix for the MZI is diagonal:
\begin{equation}\label{eq_3}
\mathrm{MZI}\left( {V_1, V_2 } \right) =
\begin{bmatrix}
   \exp \left[ j\left( \frac{V_1 }{V_\pi} \pi  + \Phi _0  \right) \right] & 0  \\
   0 & \exp \left[ j\left( \frac{V_2 }{V_\pi} \pi \right) \right]
\end{bmatrix},
\end{equation}
where $V_\pi$ is the $\pi$ phase shift voltage for each phase
modulator, and $\Phi _0$ is the zero-voltage phase between two
modulators arising from the length imbalance of two arms, denoted
$\Delta L$. The value of $\Delta L$ introduces temperature-caused
instabilities into the system and plays an important role in
determining the operating wavelength and voltages of the MZI as seen
below. To find its value, we perform a wavelength scan over a range
of more than 2~nm around 1550~nm. When a linear polarizer and a
quarter waveplate are placed after the PBC, the transmitted
intensity $I_{tr}$ exhibits an oscillation:
\begin{equation}\label{eq_Phi_0}
 I_{tr} = {1 \over 2} I_0 \left( {1 + \cos \left( {2 \vartheta } \right)\cos \Phi _0 } \right).
\end{equation}
where $\vartheta$ is the angle of polarizer with respect to the
optical axis of MZI, $I_0$ is the input intensity of MZI. $\Phi_0$
is a function of wavelength $\lambda$ as $ \Phi _0 \left( \lambda
\right) = n_1 \frac{2\pi}{\lambda}\Delta L$. Here $n_1$ is the effective
refractive index in the optical fiber, taken as constant over the
wavelengths of interest; for wider wavelength range, dispersion in the fiber must be taken into account.
If we use wave number $m = {1 \over \lambda}$, then $\Phi _0$
becomes a linear function of $m$ as $ 2 n_1 \pi m \Delta L  $.
Eq. \ref{eq_Phi_0} is plotted in Fig.~\ref{fig:delta_L} versus $m$ and compared with the experimental results. Based
on the data in Fig.~\ref{fig:delta_L}, we compute $\Delta L =
6.0\times 10^{-3}$~m. Since $n_1$ is also a function of temperature
and wavelength, exactly determining the optimal $\Phi _0$ requires
further fine tuning the input wavelength.
In order to study polarization intensities as Stokes vector
elements, we need to convert the Jones calculus into a Mueller
calculus. The Mueller matrix for a given Jones matrix can
be~\cite{Fujiwara2007Spectros} converted using $ M = A \cdot \left(
{J \otimes J^* } \right) \cdot A^{ - 1}$ with
$$
A = \begin{bmatrix}
   1 & 0 & 0 & 1  \\
   1 & 0 & 0 & -1  \\
   0 & 1 & 1 & 0  \\
   0 & j & -j & 0
\end{bmatrix}.
$$
By transforming the matrix $[J]$ in Eq.~\ref{eq_1}, we find the
Mueller matrix corresponding to our polarization modulator:
\begin{equation}\label{eq_M0} M_O  = \begin{bmatrix}
   1 & 0 & 0 & 0  \\
   0 & \cos (\Theta) \cos \left( {2\delta } \right) & \cos (\Theta)\sin \left( {2\delta } \right) & -\sin (\Theta)\\
   0 & \sin (\Theta)\cos \left( {2\delta } \right) & \sin (\Theta)\sin \left( {2\delta } \right) & \cos (\Theta)\\
   0 & \sin \left( {2\delta } \right) & -\cos \left( {2\delta } \right) & 0
  \end{bmatrix},
\end{equation}
where $\Theta = \frac{V_1  - V_2}{V_\pi}\pi  + \Phi _0$ is the
voltage modulation angle mediated by $V_1$ and $V_2$, which are the
phase modulator voltages, and $\Phi _0$ has been discussed by
Eq.~(\ref{eq_Phi_0}); $\delta$ is the angular deviation from perfect
$45^\circ$ input polarization. Now we take the input as linearly
polarized as in Eq.~(\ref{Jones_in}), such that the input Stokes
vector is $S =
\begin{pmatrix} 1 & 1 & 0 & 0  \end{pmatrix}^T$
The output of the polarization modulator (multiplication by $M_O$ of
Eq.~(\ref{eq_M0})) is then
\begin{equation}\label{eq:Stokes}
\begin{bmatrix}
   {S_0 }  \\
   {S_1 }  \\
   {S_2 }  \\
   {S_3 }
\end{bmatrix}
 = \begin{bmatrix}
   1  \\
   \cos \left( \frac{V_1  - V_2 }{V_\pi}\pi  + \Phi _0 \right)\cos \left( {2\delta } \right)  \\
   \sin \left( \frac{V_1  - V_2 }{V_\pi}\pi  + \Phi _0 \right)\cos \left( {2\delta } \right)  \\
   \sin \left( {2\delta } \right)
\end{bmatrix}.
\end{equation}

\begin{figure}
\centering
  \includegraphics[width=0.4 \textwidth]{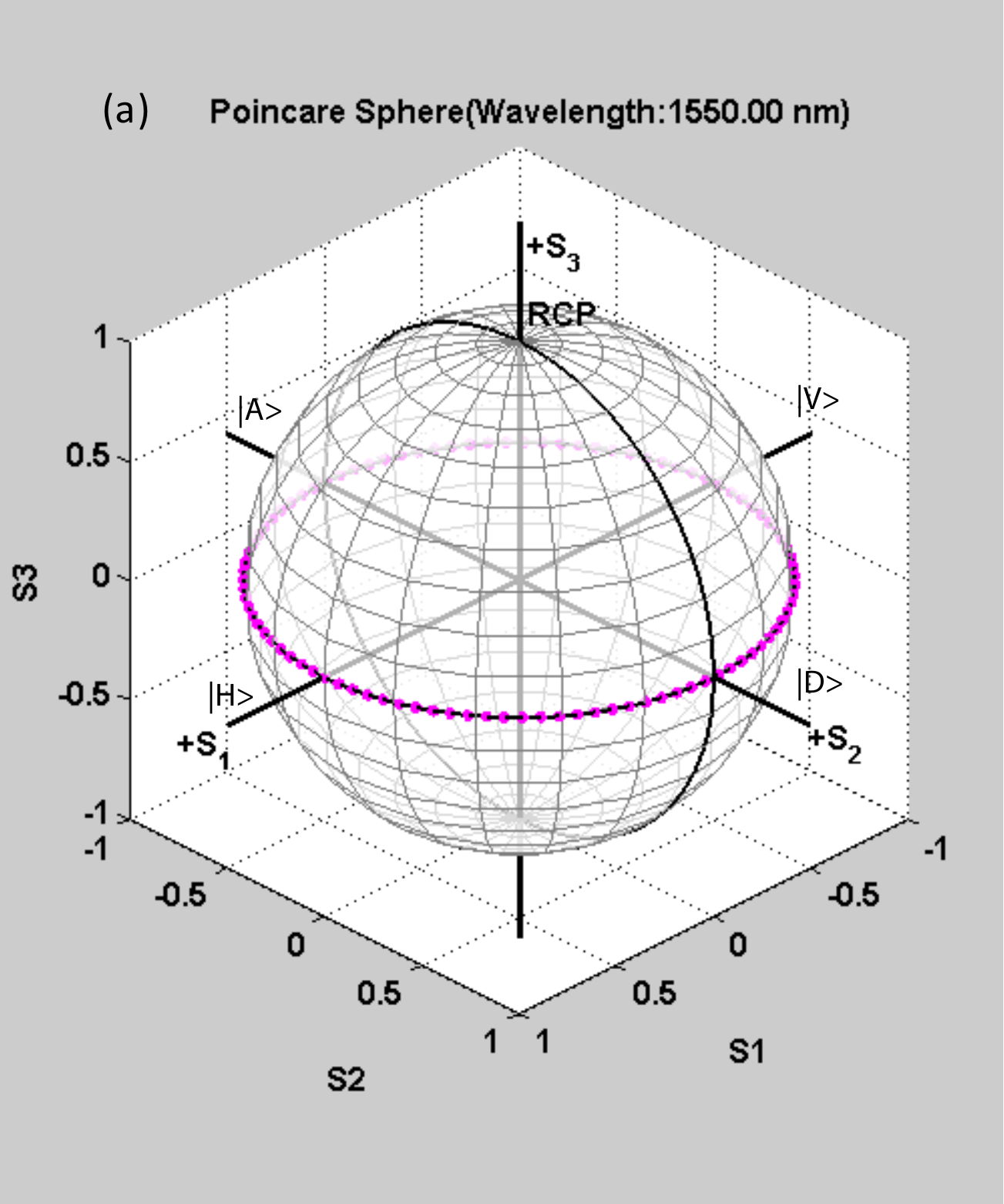}
  \includegraphics[width=0.4 \textwidth]{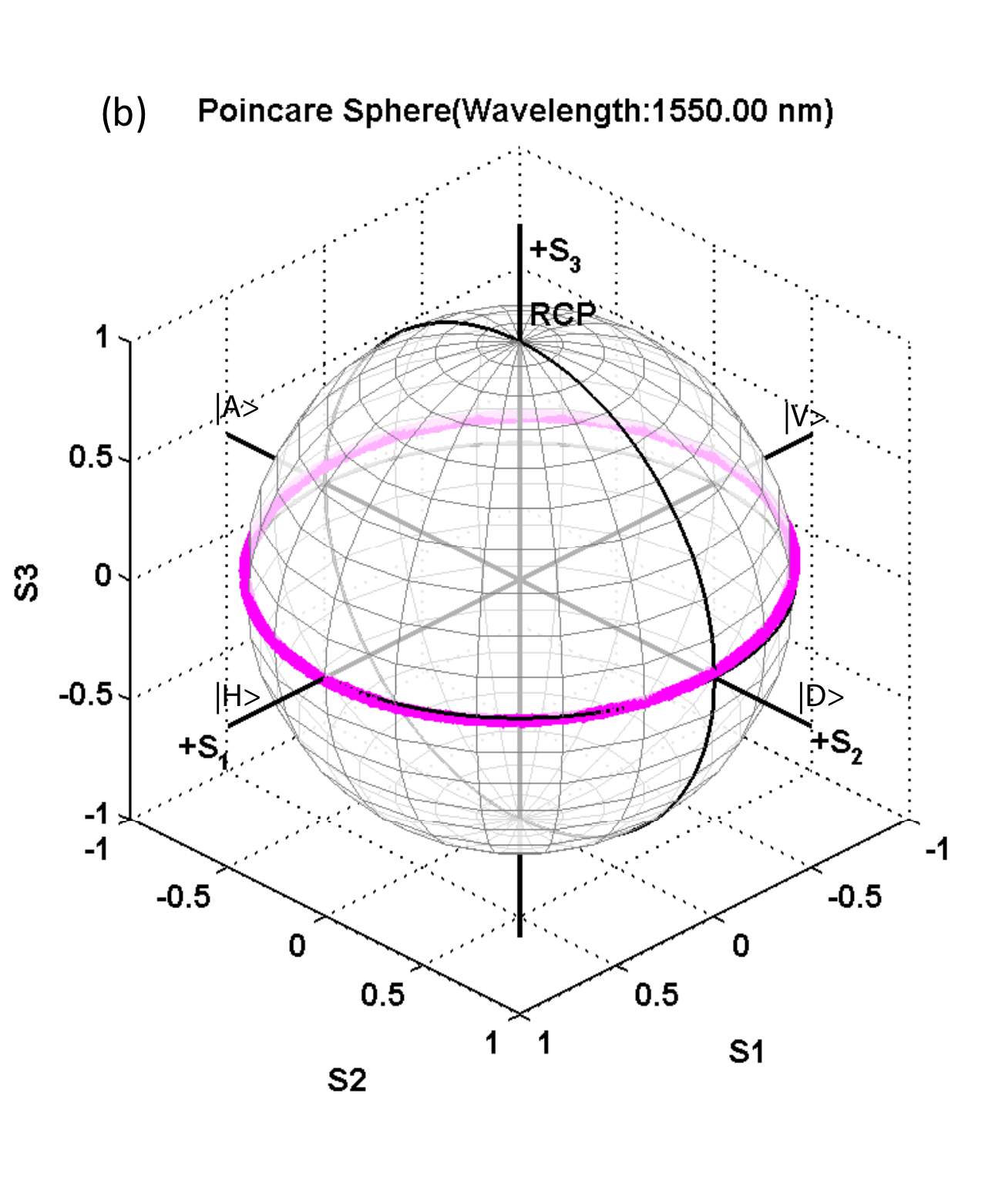}
\caption{Poincar\'{e} sphere traces of the light passing through the
polarization modulator, due to a triangular shaped voltage. (a)
Simulation results; (b) experimental measurements. (a) and  (b) are
determined after the polarization controller FPC\_2 and include the
unitary transformation $\mathrm{QWP}(\frac{\pi}{4})$ to bring the
modulation to the equator of the Poincar\'e sphere. The polarization
states are measured by an Agilent N7788B Optical Component Analyzer,
and the Polarization Navigator software
package.}\label{fig:Poincare}
\end{figure}

Eq.~(\ref{eq:Stokes}) indicates the two degrees of freedom that are
outside direct control by the phase modulator voltages: the
intensity imbalance of the arms modelled by deviation angle
$\delta$, and the initial phase angle $\Phi_0$ which stems from the
optical path difference between the MZI arms. The non-ideal
splitting angle $\delta$ can usually be eliminated by carefully
adjusting fiber polarization controller FPC\_1 in
Fig.~\ref{fig:system_architecture}. To set the initial phase
$\Phi_0$ to the optimal value ($\pi/4$) where minimal driving
voltages are needed, we varied the wavelength of the telecom laser
(HP~8168 is illustrated in Fig.~\ref{fig:system_architecture}) based
on Eq.~(\ref{eq_Phi_0}). After these two adjustments, we have the
output polarization states
\begin{equation}\label{eq:Stokes_Adjusted}
\begin{bmatrix}
   {S_0 }  \\
   {S_1 }  \\
   {S_2 }  \\
   {S_3 }
\end{bmatrix}
 = \begin{bmatrix}
   1  \\
   \sin \left( \frac{V_1  - V_2 }{V_\pi  } \pi + \frac{\pi}{4} \right)  \\
   \cos \left( \frac{V_1  - V_2 }{V_\pi  } \pi + \frac{\pi}{4} \right)  \\
   0
\end{bmatrix}.
\end{equation}
Thus, using our modulator, it is possible to output any polarization
on the equator of the Poincar\'e sphere by controlling the phase
voltages $V_1$ and $V_2$.  Simulation results are shown in
Fig.~\ref{fig:Poincare} (a). To confirm the accuracy of the
simulation, we have performed polarization measurements with an
Agilent N7788B optical component analyzer; the results of this are
shown in Fig.~\ref{fig:Poincare} (b). To control the phase
modulators for these measurements, we supply two triangular waves
with a $\pi$ phase difference to the RF ports of the modulators.

The states of interest for BB84 and corresponding ideal control
voltages are listed in Table~\ref{tb:InputOutput}. Driving voltages are minimized when $\Phi_0 = \frac{\pi}{4}$ to
reduce power dissipation in the driver. The driving voltages are chosen
such that the voltage settings are symmetric around zero volts, resulting in
zero DC bias voltage. The maximum sampling speed of the polarization
analyzer we used is 0.1~MHz, much less than the 76~MHz system clock
rate. Thus, to characterize the polarization states at higher speed,
we developed a new method which we describe below.

\begin{table}
  \centering
\caption{The driving voltage settings of two arms and corresponding output polarization states. The value
of voltages labeled in the right hand figure is $V_1 - V_2$. }\label{tb:InputOutput}
\begin{tabular}{c|c||c}
  \hline
  \hline
   $V_1$ & $V_2$  & State  \\
  \hline
  $ \frac{V_\pi}{8}$ & $ \frac{- V_\pi}{8} $ & H   \\
  $ \frac{-V_\pi}{8}$ & $ \frac{V_\pi}{8}$   & D  \\
  $ \frac{-3V_\pi}{8}$ & $ \frac{3V_\pi}{8}$ & V   \\
   $ \frac{3V_\pi}{8}$ &  $ \frac{-3V_\pi}{8}$ & A   \\
  \hline
  \hline
\end{tabular}
\begin{tabular}{c}
      \includegraphics[width=0.4 \textwidth]{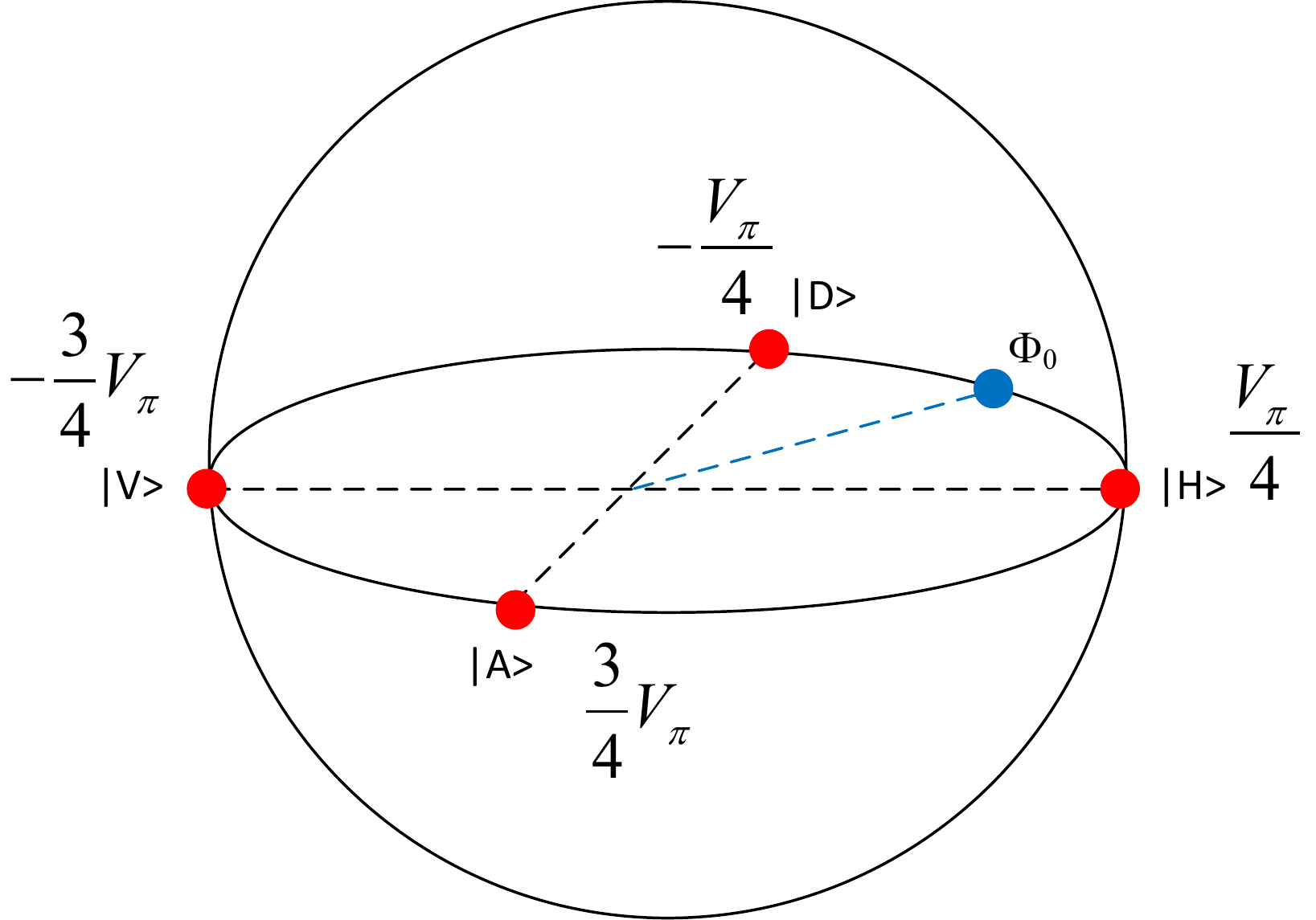}\\
\end{tabular}
\end{table}
\subsection{Polarization state characterization}
We wish to know the exact output polarization state described by
Eq.~(\ref{eq:Stokes}) at full modulation speed. It is impossible to
do a direct characterization of the Stokes vectors using standard
polarimeters due to the high repetition rate of the modulator. Here we use a quarter waveplate (QWP) followed by a linear polarizer with a repeated
modulation sequence that allows us to extract the Stokes parameters
of each state in the sequence. The intensity $S_j^\pm$ after the QWP
and polarizer is calculated in Ref.~\cite{Berry_AppOpt_77}
\begin{align}\label{eq:PolMeasure}
S_j^\pm  = & \frac{1}{2}\big\{ S_0  + \left( {S_1 \cos 2\beta_j  +
S_2 \sin 2\beta_j } \right)\cos 2\left(
{\alpha_j  - \beta_j } \right)  \nonumber \\
 \quad + & \left[ {\left( {S_2 \cos 2\beta_j  - S_1 \sin 2\beta_j }
\right)\cos \Delta  + S_3 \sin \Delta } \right] \nonumber \\
& \times \sin 2\left( {\alpha_j  - \beta_j } \right) \big\},
\end{align}
where $\alpha_j$ is the angle of the linear polarizer's transmission
axis and $\beta_j$ is the angle of the fast axis of the QWP---in our
experiment, angles $\alpha_j$ and $\beta_j$ are both with respect to
horizontal. The retarder phase delay $\Delta$ is $\pi/2$ for a QWP,
however in real devices there exists non-trivial deviations to this
ideal value. In our setup, we use a QWP from Thorlabs and find $7\%$
offset from $\frac{\pi}{2}$.

\begin{figure}
\centering
  \includegraphics[width= 0.45 \textwidth]{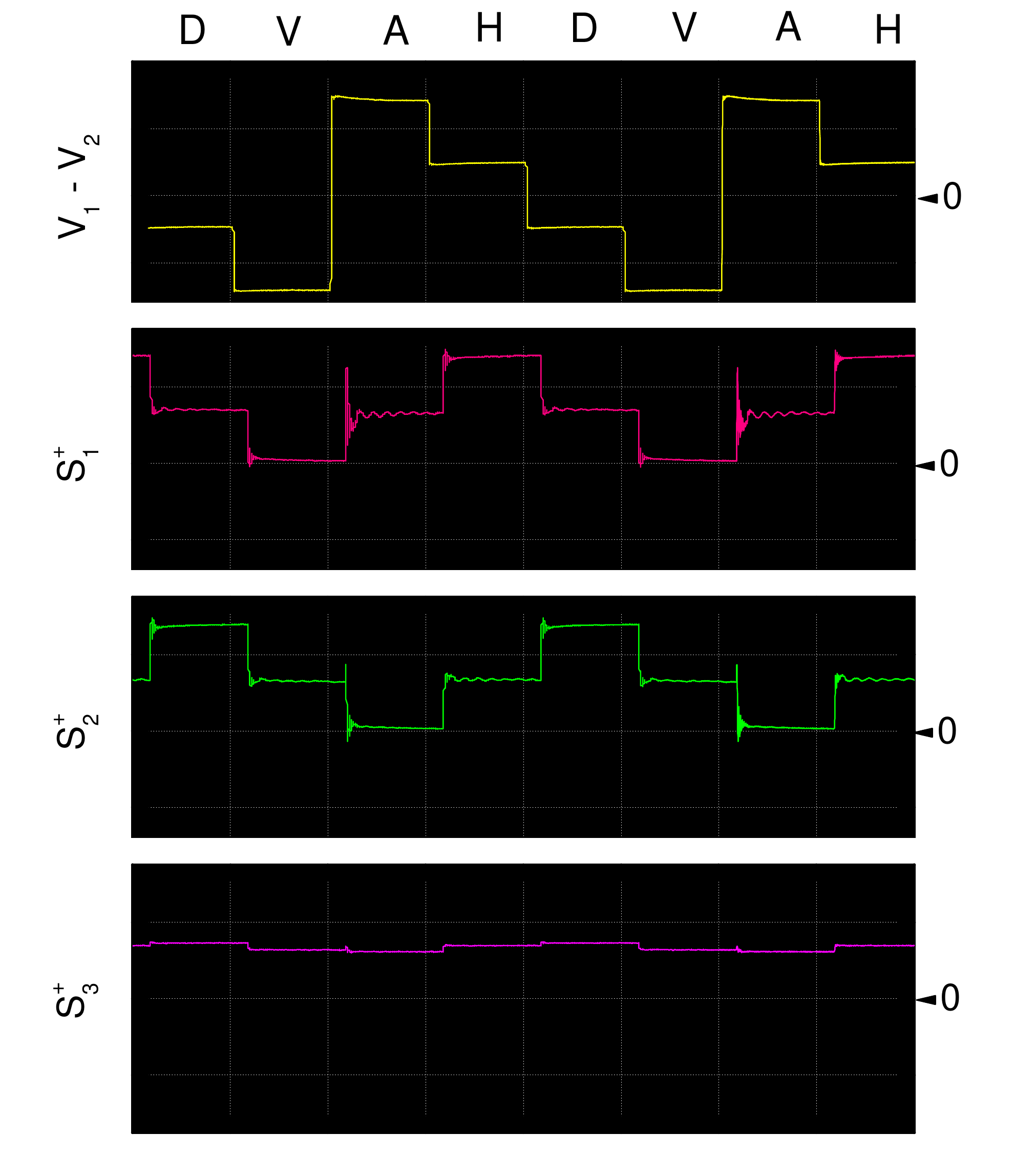}
    \includegraphics[width= 0.4 \textwidth]{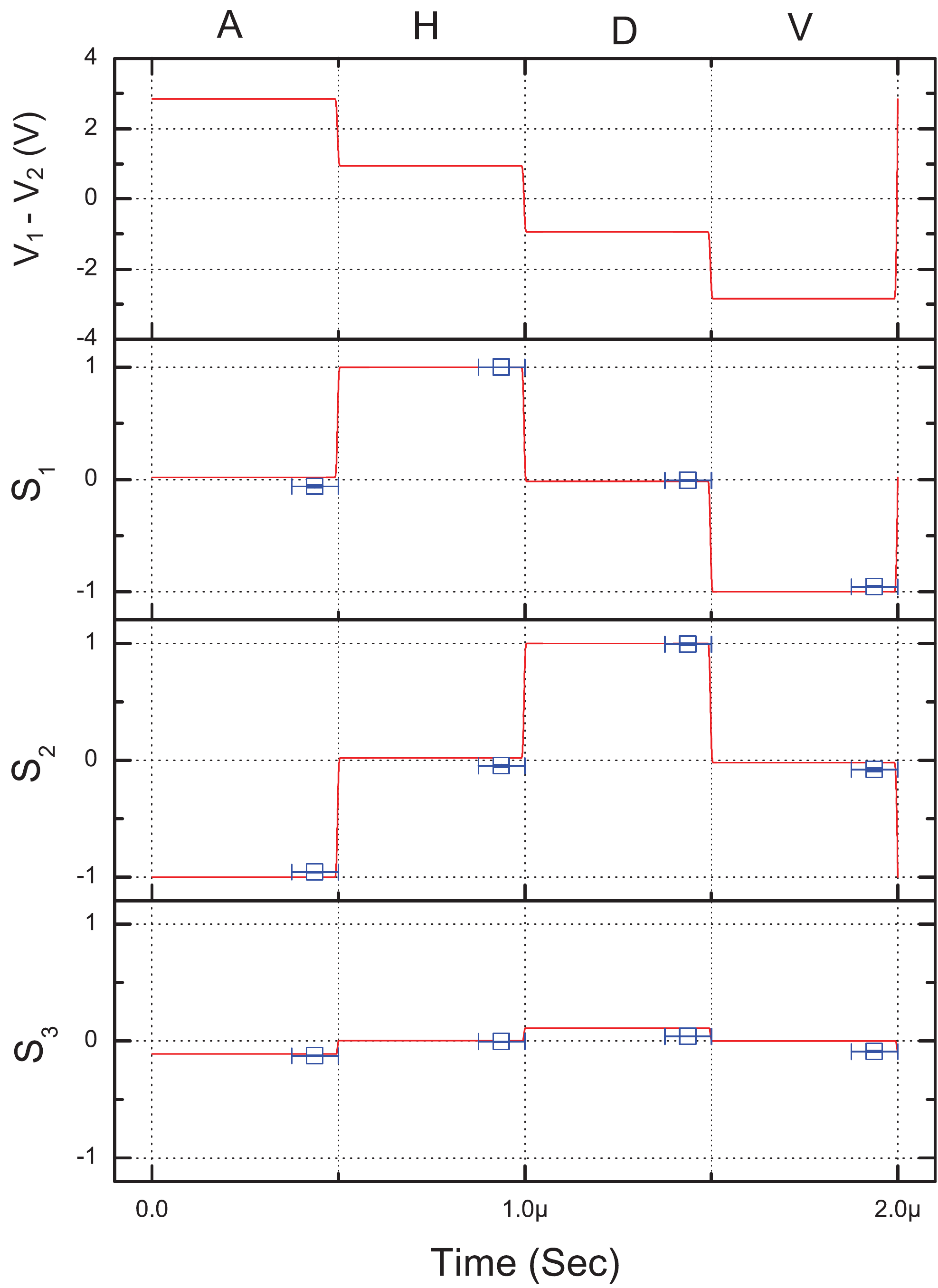}
\caption{(a) The oscilloscope traces to measure Stokes vector
including differential voltage $V_1 - V_2$ between two phase
modulators located at two arms of MZI; three measurements for
$S_1^+$, $S_2^+$ and $S_3^+$ at a wavelength of 1550.5~nm. All of
the raw signal traces were collected by an Agilent DSO8104A
Infiniium Oscilloscope. Each horizontal division was set to be 500
ns in the plot; the differential voltage plot had 2 volts per
division for its vertical trace; while the rest three plots had 100
mV per division. (b) The simulation results for wavelength $\lambda
= 1550.5$~nm, with measured three Stokes vector elements normalized
to $S_0$ (in blue hollow squares with error bars). The simulation
includes the imperfect retardance ($7 \%$) of the QWP; the time
delay between differential voltage to the polarization states has
not been taken into account.
}\label{fig:stokes_simulation_experiment}
\end{figure}

At least three measurements with the settings defined in
Table~\ref{tb:Stokes} were performed to obtain three components
($S_1^+$, $S_2^+$, and $S_3^+$) taken at 2~MHz modulation speed. The
results are displayed in
Fig.~\ref{fig:stokes_simulation_experiment}~(a). This
characterization scheme utilized an optical detector (Thorlabs
PDA10CF) which had a bandwidth of 150~MHz.

In Fig.~\ref{fig:stokes_simulation_experiment}~(a), there is a time
displacement of around 100 ns between the differential voltage $V_1
- V_2$ and polarization states due to the delays between the optical
and electrical signals. The spike overshoot is due to the finite
transition time across multiple polarization states, when the
differential driving voltage has the largest step. The transition
ring ripples after the spike is caused by the limited response time
of the optical PIN diode, but the actual polarization states should
not be affected. Moreover, since the upconversion SFG process only
occurs in a time overlap (less than 1~ns) when both telecom and 810
nm photons arrive, the polarization states of telecom photons within
a few percentage of each repetition cycle are responsible for the
green photon states. We are able to completely avoid the transition
spike by moving the overlap time to the end of each cycle.

In the subsequent simulation, we have modeled the entire IPM system
as well as the polarization characterization optics. The
imperfection of QWP was also take this into account. The degree of
polarization (DOP) was experimentally found to be over $99 \%$,
hence $S_0$ can be determined by the averaged maximum intensity. We
then use the relations $S_1 = 2 S_1^+ - S_0$, $S_2 = 2 S_2^+ - S_0
$, and $S_3 =2 S_3^+ - S_0 $ to determine the exact polarization
states in H/V, D/A, and L/R bases, respectively, for each modulation
setting ($V_1$ and $V_2$). The simulation results compared with the
extracted Stokes vector elements $S_1$, $S_2$ and $S_3$ (in hollow
square) are displayed in
Fig.~\ref{fig:stokes_simulation_experiment}~(b). For each extraction
point, the $x$ error bar indicates the length of time (125~ns) to
average the polarization states; the $y$ error bar indicates the
uncertainty of each extraction.

\begin{table}
  \centering
\caption{Measurement settings for polarization state characterization}\label{tb:Stokes}
  \begin{tabular}{c|c|c}
    \hline
    \hline
       $S_j^\pm$      & Linear Polarizer ($\alpha_j$) & Quarter-Wave Plate ($\beta_j$) \\
    \hline
     $S_1^+$ & 0$^\circ$ & 0$^\circ$ \\
     $S_2^+$ & 45$^\circ$ & 45$^\circ$ \\
     $S_3^+$ & 45$^\circ$ & 0$^\circ$ \\
    \hline
    \hline
  \end{tabular}
\end{table}

\subsection{Temperature Stability of the Modulator}
The temperature stability of the polarization state plays a critical
role for long-term operation of this device. To measure the
performance of a given polarization state against the variation of
ambient temperature, we use a constant optical power and wavelength
at the input port of our polarization controller, and measure the
output power of four polarization components by our
in-house-developed polarimetry method described in the previous
section.

\begin{figure}
\centering
\includegraphics[width=0.45 \textwidth]{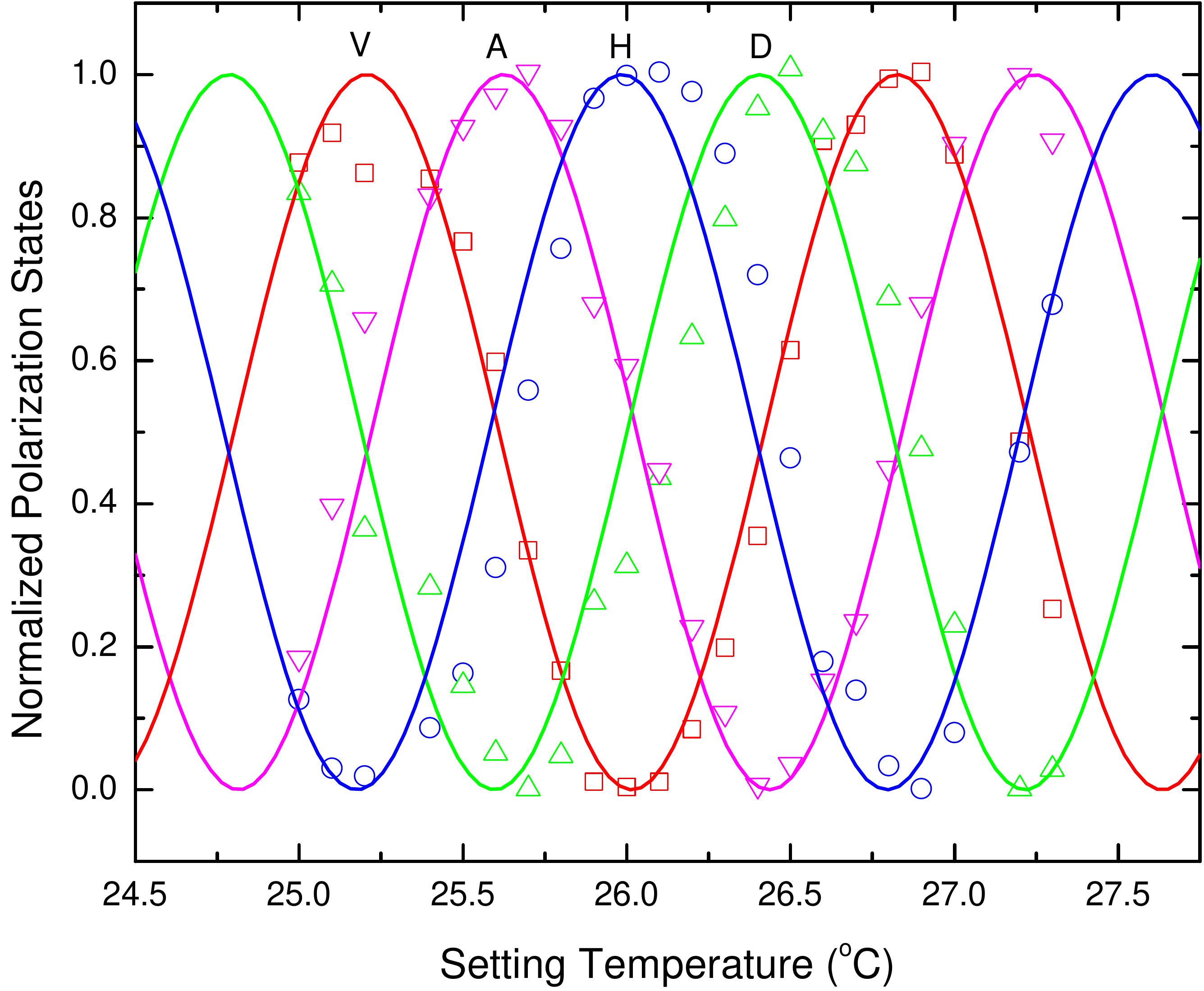}
\caption{The measured power of four polarization states as
temperature varies. The setting of temperature is monitored and
maintained by a Thorlabs TC-200 temperature controller driving a
point contact heater in the enclosure of the MZI polarization
controller.}\label{fig:wcp_temp}
\end{figure}

Fig.~\ref{fig:wcp_temp} shows the measurement results for H/V and
D/A components. The temperature is maintained by a Thorlabs TC-200
controller. In this measurement, no voltage is applied to either the
phase or intensity modulators, but the initial phase $\Phi_0$ in the
MZI becomes a function of temperature. By carefully eliminating the
L/R components, the output polarization state sweeps across the
equator of the Poincar\'{e} sphere (provided that the temperature
variation is big enough) as indicated by Fig.~\ref{fig:wcp_temp}.

However, at higher temperature settings when the temperature setting
is $> 30 \mathrm{^o C}$, the temperature actuator (heater) cycles
through on and off states more frequently, resulting in strong
fluctuations and poor visibility. This is due to the simple point
contact heater used---it takes some time for the temperature of the
arms of the MZI to stabilize, and more importantly two arms are not
heated up equally leading to adversely increasing the imbalance of
MZI. A two-stage heater would be desirable for achieving better long
term temperature stabilization.

\section{Experimental Results of Intensity and Polarization Modulator Implementing BB84 Protocol with Decoy States}
\subsection{Theory of BB84 protocol with decoy states}
WCP QKD, attractive for its simple designs and high rates, is
subject to the photon number splitting
attack~\cite{PhysRevLett.85.1330} due to the nonzero probability of
producing multiple photons in a single laser pulse. To securely
achieve any reasonable transmission distance requires the inclusion
of decoy states, i.e.\ pulses with differing intensities used to
bound the eavesdropper's information from multi-photon
events~\cite{PhysRevLett.91.057901,Lo_PRL_05}. We implement a
polarization-encoded protocol with vacuum+weak decoy
states~\cite{Ma_PRA_05}. To calculate the final secure key rate, we
must estimate the single-photon gain and error rate. We calculate a
lower bound of single photon gain $Q_1^L$  from
Ref.~\cite{Ma_PRA_05} as
\begin{equation}\label{eq:Q1L}
Q_1^L  = \frac{\mu ^2 \mathrm{e}^{ - \mu } }{\mu \nu  - \nu ^2
}\left( Q_\nu  \mathrm{e}^\nu   - Q_\mu  \mathrm{e}^\mu  \frac{\nu
^2 }{\mu ^2 } - \frac{\mu ^2  - \nu ^2 }{\mu ^2 }Y_0  \right),
\end{equation}
where $\mu$ is the average photon number for signal states and $\nu$
for decoy states, and $Q_{\mu/\nu}$ are the gains for these two
states. $Y_0$, the vacuum yield, is determined by the detector dark
counts and background noise, and was measured between non-vacuum
pulses. Similarly, the upper bound on the error rate,
$\varepsilon_1^U$, of single photon states is
\begin{equation}\label{eq:e1U}
\varepsilon_1^U  = \frac{E_\nu  Q_\nu \mathrm{e}^{\nu} - e_0 Y_0
}{\nu Q_1^L }\mu \mathrm{e}^{ - \mu },
\end{equation}
where $E_\nu$ is the total error rate for decoy states, and $e_0 = 0.5$ is the vacuum error rate. After
evaluation of the parameters in Eqs.~(\ref{eq:Q1L}) and~(\ref{eq:e1U}), we can find the lower bound of the
secret key rate per pulse, $R$:
\begin{equation}\label{eq:SecureRate}
R = q L_{\mu (\nu^\prime)} \left\{ { - Q_\mu  f\left( \mathrm{QBER}
\right) H_2 \left( \mathrm{QBER} \right) + Q_1^L \left[ {1 - H_2
\left( {\varepsilon_1^U } \right)} \right]} \right\}.
\end{equation}
where $q = 1/2$ is the basis reconciliation factor, $L_{\mu \nu} =
N_\mu/(N_\mu + N_\nu)$ is the ratio of signal pulses to the total
pulses, and $\mathrm{QBER}$ is the total quantum bit error rate for
signal pulses. For the calculations here, a constant error
correction efficiency $f(\mathrm{QBER})$ of 1.22 was assumed.
\begin{figure}
\centering
  \includegraphics[width= 0.4 \textwidth]{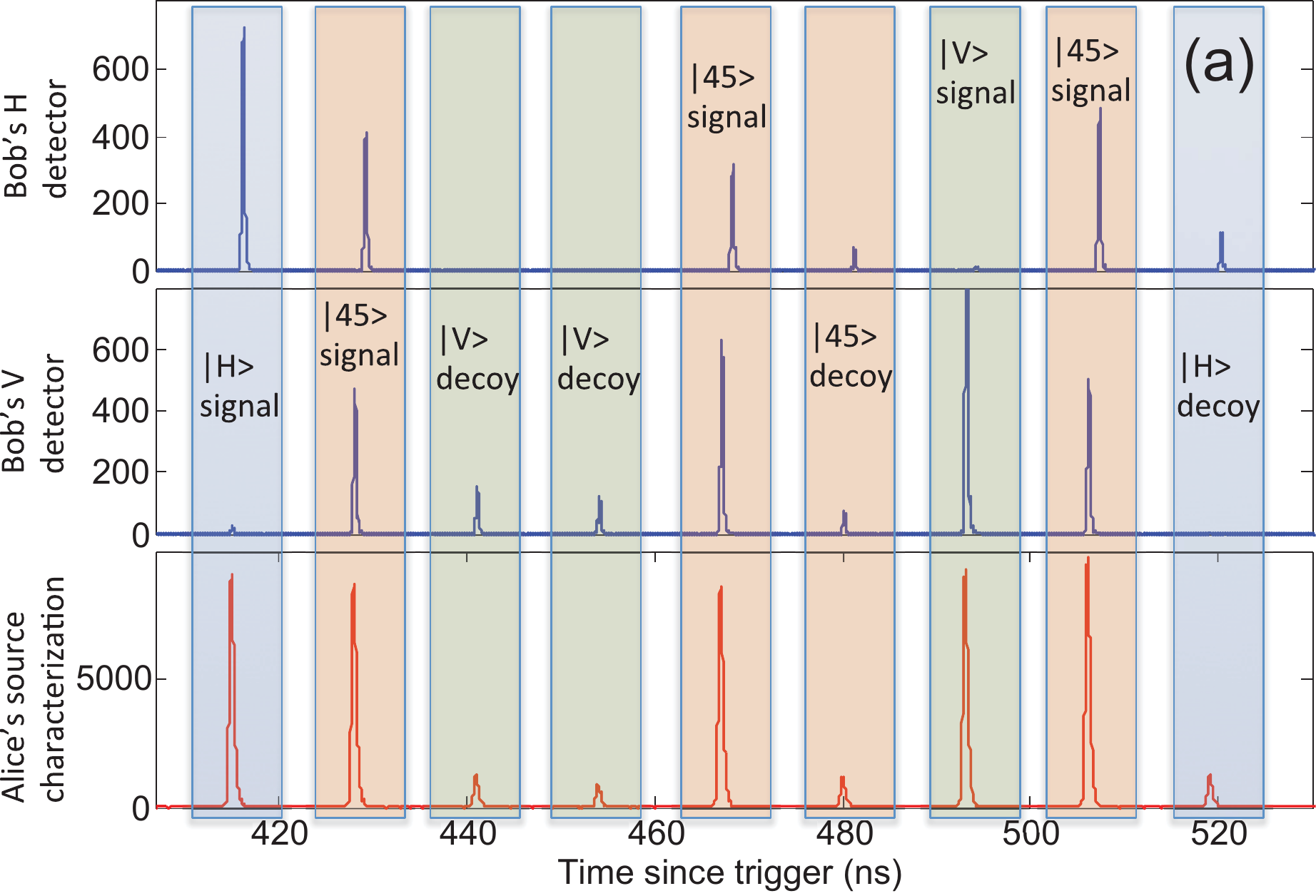}
  \includegraphics[width= 0.4 \textwidth]{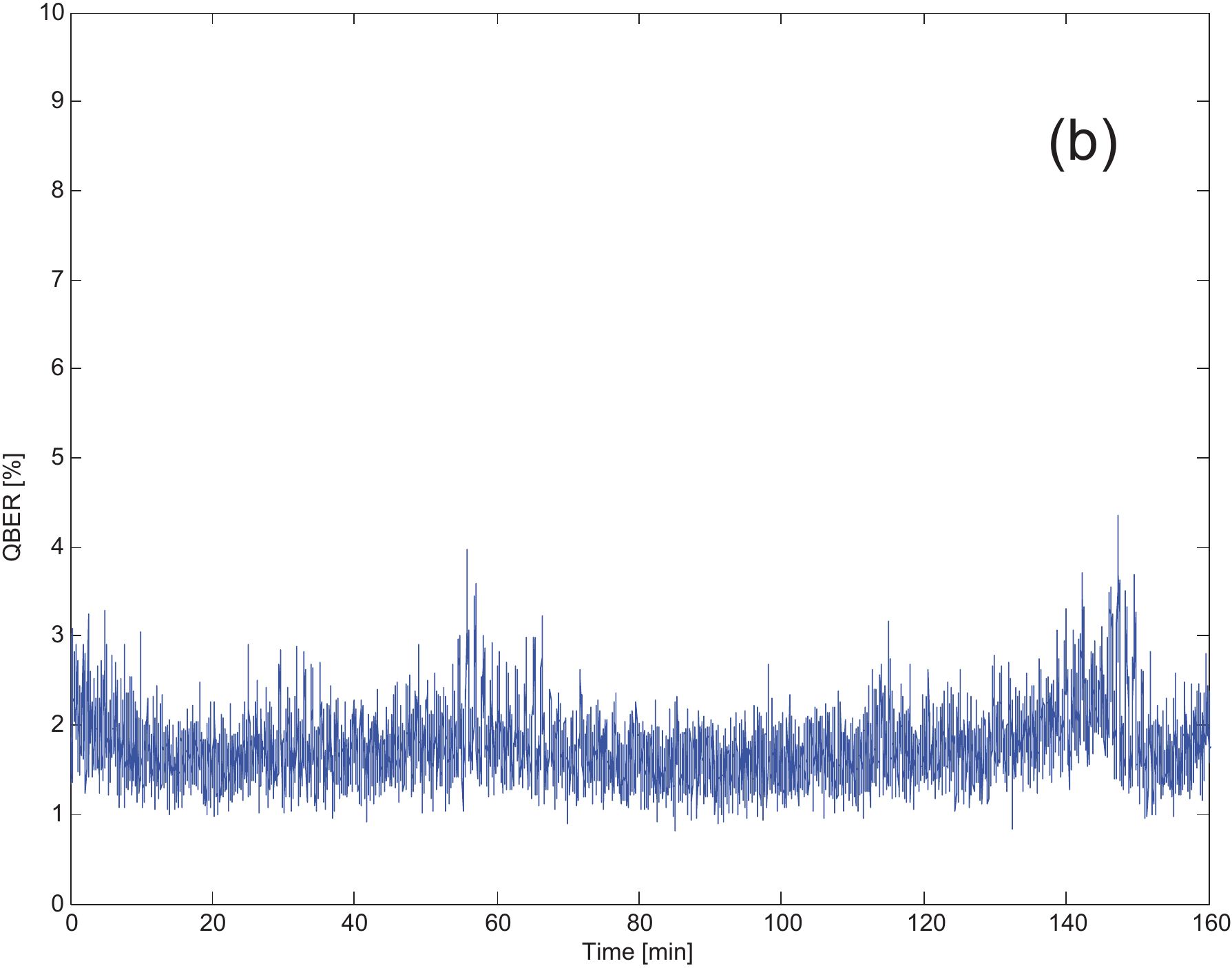}
\caption{(a) Experimental measurements for the decoy+signal QKD
states accumulated in the histogram format
(see~\cite{Meyer-Scott_thesis_master} in which Bob's detector was
fixed to an H/V basis measurement.). $|45\rangle$ indicates either a
D or A polarization state which is not resolved by our two-detector
receiver in this basis. (b) Stability measurement of the overall QKD
system-wide QBER over consecutive 160 minutes at 25~dB total loss,
showing the averaged QBER at $1.8\pm0.9\%$. The system clock rate is
76~MHz triggered from the Ti-Sapphire mode-locked
laser.}\label{fig:histogram}
\end{figure}
\subsection{Experimental Demonstration of High Loss QKD System}
\begin{figure}
\centering
  \includegraphics[width= 0.5 \textwidth]{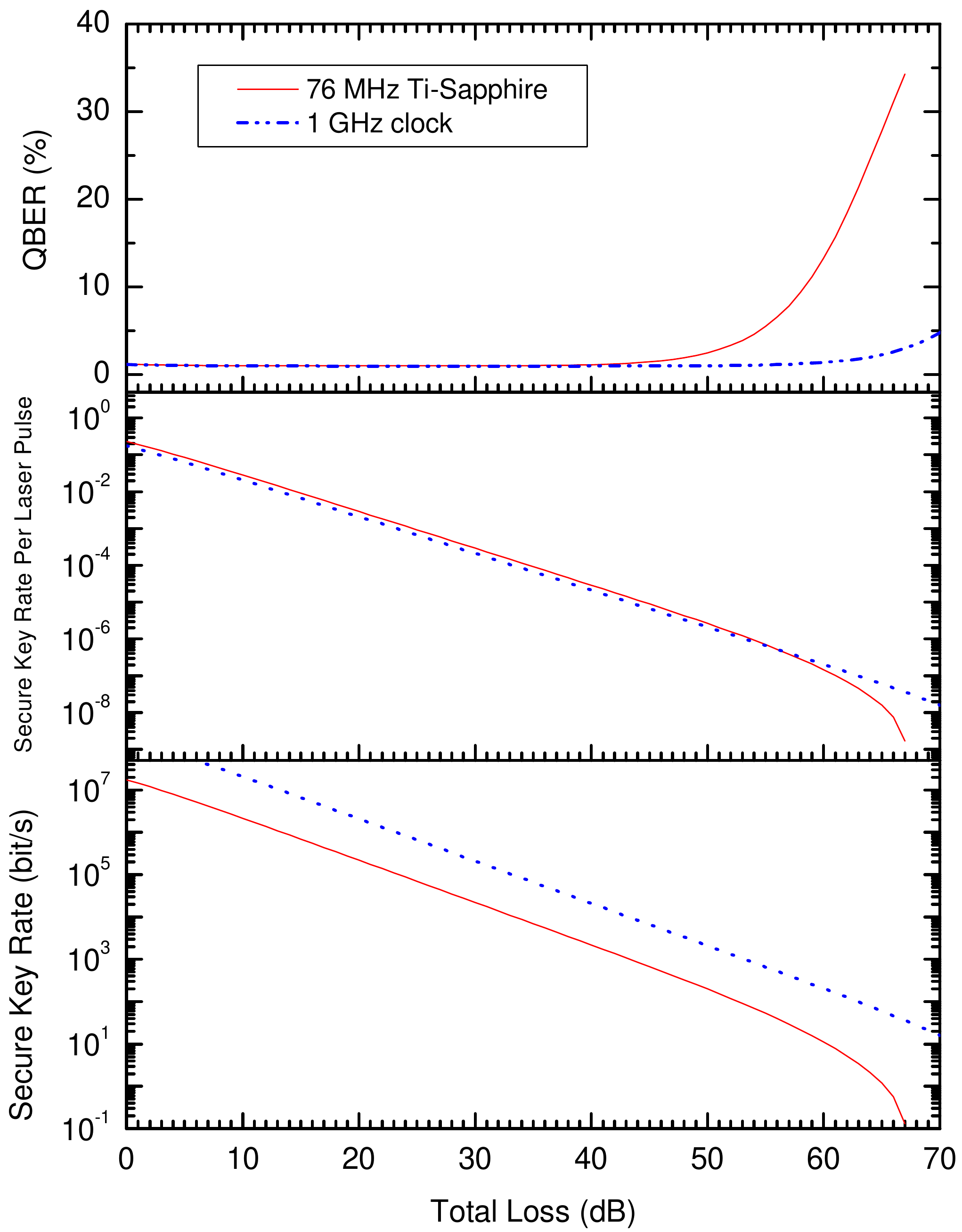}
\caption{The projected performance of our decoy state BB84 QKD
system. We show the simulation results of QBER, secure key rate
probability per pumping laser pulse, and the final key rate under
total channel loss. We assume two pumping laser rates: the currently
used Ti-Sapphire laser pumping rate of 76~MHz (solid red line), and
the projected 1~GHz clock rate (dashed blue
line).}\label{fig:decoy_qkd_sim}
\end{figure}
Our complete system exhibits high performance: when the 810~nm laser
is not mode-locked (i.e.\ continuous wave), the output photons at
532~nm have a fidelity $>$99\% with the desired polarization states.
Upon mode-locking and fast modulation, we find that our system
fidelity including all losses and error sources is maintained at
$>$98\% in the H/V polarization basis, and $>$95\% in the D/A basis.
The modulation achieves accurate average photon number per pulse for
both decoy and signal states, sufficient to perform successful QKD.
Fig.~\ref{fig:histogram}~(a) demonstrates the measured photon output
for signal and decoy states with various polarizations. We
additionally implemented an automated polarization alignment
procedure in the quantum receiver to keep high fidelity for all four
states, the details of which will be published elsewhere.
Fig.~\ref{fig:histogram}~(b) illustrates a QBER stability
measurement over more than 160 minutes with ambient temperature
fluctuation ${<}0.5^\circ$C. The QBER shows an average value of
$1.8\pm0.9\%$, a very small variation. Our simulations of
ground-to-space quantum channels show that the overall QBER of our
system is sufficient for a satellite
uplink~\cite{Link_analysis_paper}.

Finally, we calculate the secure key rate based on Eq.~(\ref{eq:SecureRate}). The result is shown in
Fig.~\ref{fig:decoy_qkd_sim}, in which the detector efficiency is 60\% and the dark count rate is 50~counts/s.
The intrinsic QBER is 1\%, which accounts for the optical misalignments in the telecom modulator, SFG
source, and receiver. In practice, stray light will lead  to slightly higher measured QBER than the
simulated value under the same channel loss. The signal and decoy states are 0.6 and 0.2 photons per pulse
respectively. QKD beyond 60~dB total loss is possible with this system.

\section{Conclusion and Outlook}
We have presented our design and implementation of a Mach-Zehnder
configuration, all-fiber, intensity and polarization modulator. This
design features very high switching speed and high polarization
visibility. Our modulator serves as a secret key encoder which can
easily be upgraded to GHz speeds owing to the rich choice of
photonic devices for telecom wavelengths. We used this modulator and
the SFG process to produce 532~nm green photons at a repetition rate
near one hundred MHz, which is the highest speed for this operating
wavelength demonstrated so far. The green photons show excellent
polarization state fidelity, and are compatible with the highest
figure of merit single photon detectors commercially available.

The quantum optical simulations in Fig. \ref{fig:decoy_qkd_sim} show QKD is possible over channels beyond
60~dB of loss. Additionally, we measured the raw key rates, average photon numbers for signal and decoy
states, and quantum bit error rates with high channel losses. The performance fulfills the design goals of
the telecom modulator and agrees with our quantum optics simulation results. Assuming an average loss of
45~dB for a LEO satellite passage, the final secure key rate is 155~bit/s.

If we employ a 1~GHz repetition rate pumping laser, the operating
clock rate of our telecom modulator can be readily boosted
accordingly, thanks to the technological advancement of
FPGA~\cite{GHzFPGA} and RF power electronics~\cite{GHzRFDriver}.
Assuming this increased speed, the QBER, secure key rate per pulse
and per second are displayed in Fig.~\ref{fig:decoy_qkd_sim} as
dashed lines. At the same average LEO satellite channel loss, the
secure key rate will be ${\sim}2300$~bit/s, an increase of more than
an order of magnitude, and better than the repetition rate increase
factor owing to an improved signal-to-noise ratio. Consequently, we
present our system that is able to overcome the high channel loss
that plagues proposals for a satellite uplink QKD system with our
current setup. Our experimental results verify the feasibility of
using such a source in a near term satellite mission. Using higher
speed modulation such as 1 GHz rate, it is possible to achieve
higher secrete key rate and to withstand higher channel losses.

\section{Acknowledgment}
This work is supported by Natural Sciences and Engineering Research Council of Canada (NSERC), Canadian
Space Agency (CSA), COM DEV International Ltd., Canada Foundation for Innovation (CFI), Industry Canada,
and Canadian Institute for Advanced Research (CIFAR). We would like to thank Agilent Technologies Inc.,
and General Photonics Inc. for providing us optical measurement instruments.
%
\end{document}